\documentclass{elsart}
\usepackage{amsmath}
\usepackage{color}
\usepackage{amssymb}
\usepackage{graphicx}
\usepackage{rotating}
\usepackage{setspace}
\usepackage{url}

\newcommand \be{\begin{equation}}
\newcommand \ba{\begin{eqnarray}}
\newcommand \ee{\end{equation}}
\newcommand \ea{\end{eqnarray}}

\begin{document}

\begin{frontmatter}

\title{Fundamental Factors versus Herding in the 2000-2005
US Stock Market and Prediction}

\author[ecust]{\small{Wei-Xing Zhou}},
\author[UCLA,nice]{\small{Didier Sornette}\thanksref{ADDR}}
\address[ecust]{State Key Laboratory of Chemical Reaction
Engineering, East China University of Science and Technology,
Shanghai 200237, China}
\address[UCLA]{Institute of Geophysics and Planetary Physics and
Department of Earth and Space Sciences, University of
California, Los Angeles, CA 90095}
\address[nice]{Laboratoire de Physique de la Mati\`ere Condens\'ee,
CNRS UMR 6622 and Universit\'e de Nice-Sophia Antipolis, 06108
Nice Cedex 2, France}
\thanks[ADDR]{Corresponding author. Department of Earth and Space
Sciences and Institute of Geophysics and Planetary Physics,
University of California, Los Angeles, CA 90095-1567, USA. Tel:
+1-310-825-2863; Fax: +1-310-206-3051. {\it E-mail address:}\/
sornette@moho.ess.ucla.edu (D. Sornette)\\
\url{http://www.ess.ucla.edu/faculty/sornette/}}

\begin{abstract}

We present a general methodology to incorporate fundamental economic
factors to our previous theory of herding to describe bubbles and
antibubbles. We start from the strong form of Rational Expectation
and derive the general method to incorporate factors in addition to
the log-periodic power law (LPPL) signature of herding
developed in ours and others' works. These
factors include interest rate, interest spread, historical
volatility, implied volatility and exchange rates. Standard
statistical AIC and Wilks tests allow us to compare the
explanatory power of the different
proposed factor models. We find that the historical volatility
played the key role before August of 2002. Around October 2002, the
interest rate dominated. In the first six months of 2003, the
foreign exchange rate became the key factor. Since the end of 2003,
all factors have played an increasingly large role. However, the
most surprising result is that the best model is the second-order
LPPL without any factor. We thus present a scenario for the future
evolution of the US stock market based on the extrapolation of the
fit of the second-order LPPL formula, which suggests that herding is
still the dominating force and that the unraveling of the US stock
market antibubble since 2000 is still qualitatively similar to
(but quantitatively different from) the Japanese Nikkei case after 1990.

\end{abstract}


\end{frontmatter}

\section{Introduction}

As an extension of \cite{SZ2002}, since October 2002, we have posted
an analysis of the US (and later other) markets after the collapse
of the ``information technology bubble,'' forecasting a continuation
of the downward correction, decorated with ups and downs organized
according to a pattern we call ``log-periodicity.'' Monthly
updates\footnote{See the website at
$http://www.ess.ucla.edu/faculty/sornette/prediction/index.asp\#prediction$.}
are available from October 31, 2002 till November 17, 2004.

Our projections have been based on the generalization to ``bearish''
markets (we call them ``antibubbles'' when they follow market bubble
peaks) of the concept of positive feedback based on imitation and
herding that our team has demonstrated mostly for bubbles preceding
financial crashes (see \cite{crashbook} for a review and references therein).

In December 2004, we decided to discontinue the update, concluding
that, after more than two years, our projections for the US market
have not been verified, while they have been good from the vantage
of foreigners as well as for other foreign developed markets. In
December 2004, we attributed our failure to forecast the rebound of
the market in 2003 to the fact that we had neglected all factors
except the imitation and herding behavior of investors. In
particular, our approach had neglected the fundamental input of the
economy as well as the external manipulations by the Fed and other
institutions, the impact of foreign policy and foreign investors,
the dollar effect (all these being possibly inter-related). Our
projections had taken the extreme view that all these actions are
endogenously determined and driven by the collective action of the
investors. We thought this is too simplistic because, in general,
one can understand the evolution of system as the result of an
entangled combination of endogenous organization but also as a
response to external news and exogenous shocks (see \cite{endoexo}
for a recent quantitative discussion on the issue of endogeneity
versus exogeneity). In December 2004, we concluded that our live
experiment from Oct. 2002 until Nov. 2004 had clearly demonstrated
the limits of this approximation.

We announced at the web site that we had found a methodology to
incorporate the Feds action and the dollar effect in an extended
version of our theory, to combine our herding theory with all these
other factors. The first purpose of the present paper is to present
this methodology. We start from the strong form of Rational
Expectation and derive the general formalism to incorporate factors
in addition to the log-periodic power law (LPPL) signature of
herding. Then, our first tests show that rational expectation should
be partially relaxed and we present several models combining LPPL
with different factors (interest rate, interest spread, historical
volatility, implied volatility, exchange rates). One particularly
interesting theoretical result is that this formalism allows us to
justify that the exchange rate factor is equivalent for a certain
value of parameters to take the view of foreigners as empirically
discovered in our monthly updates. We also present several standard
statistical AIC and Wilks tests to compare the different proposed
factor models.

The most surprising result is in fine that we have to prefer an
extension of the LPPL formula, the so-called second-order LPPL first
introduced in \cite{SJ97PhysA}, to
all other possible models including economic factors. We thus present
a scenario for the future evolution of the US stock market based
on the extrapolation of the fit of the second-order LPPL formula.

\section{Model 1: Rational Expectation bubble model with a single
volatility factor and log-periodicity}

\subsection{The stochastic discount factor}

The theory of stochastic pricing kernel \cite{Cochrane-2001}
provides a unified framework for pricing consistently all assets
under the premise that the price of an asset is equal
to the expected discounted payoff,
while capturing the macro-economic risks underlying each
security's value. This theory amounts to postulating the
existence of a stochastic discount factor (SDF) $M$ that allows
one to price all assets consistently. The SDF is also termed the pricing
kernel, the pricing operator, or the state price density and can
be thought of as the nominal, inter-temporal, marginal rate of
substitution for consumption of a representative agent in an
exchange economy. Under an adequate definition of the space of
admissible trading strategies, the no-arbitrage
condition translates into the condition that the product of the SDF with the
value process $p(t)$, of any admissible self-financing trading strategy
implemented by trading on a financial asset, must be a martingale:
\be
M(t) p(t)= {\rm E}_t \left[p(t') M(t') \right]~,
\label{martcond}
\ee
where $t'$ refers to a future date. The no-arbitrage condition
(\ref{martcond}) expresses that $p M$ is a martingale for any
admissible price $p$. Technically, this amounts simply to imposing that
the drift of $p M$ is zero.

Let us assume as a first example the following simple dynamics for
the SDF, which will lead to Model 1 for the price dynamics: \be
\frac{dM(t)}{M(t)}  = -r(t)~ dt  - \phi(t)~dW(t) + g(t) d{\hat W}~.
\label{sdf} \ee The drift $-r(t)$ of $M$ is justified by the
well-known martingale condition on the product of the bank account
and the SDF \cite{Cochrane-2001}. Intuitively, it retrieves the
usual simple form of exponential discount with time. The process
$\phi$ denotes the market price of risk, as measured by the
covariance of asset returns with the SDF. Stated differently, $\phi$
is the excess return over the spot interest rate that assets must
earn per unit of covariance with $W$, hence the negative sign in
front of $\phi(t)~dW(t)$ in (\ref{sdf}) which leads to a positive
contribution to the return $\mu$ as we now show. The last term $g(t)
d{\hat W}$ embodies all other stochastic factors acting on the SDF,
which are orthogonal to (uncorrelated with) the stochastic process
$dW$.

For the sake of pedagogy and clarity, let us provide a
straightforward derivation of Ito's calculus applied to the
condition that the process $p M$ is a martingale. For this, we form
the expectation of \be \frac{p(t+dt) M(t+dt) - p(t) M(t)}{ p(t)
M(t)} \ee whose drift (the term proportional to $dt$) must be zero
for the no-arbitrage condition to hold. This reads \ba {\rm
E}\left[\frac{p(t+dt) M(t+dt) - p(t) M(t) }{ p(t) M(t)}\right] &=&
{\rm E}\left[\frac{(p(t) + dp) (M(t)+dM) - p(t) M(t) }{ p(t) M(t)}
\right] \nonumber \\
&=& {\rm E}\left[\frac{p(t) dM + M(t) dp + dM dp }{ p(t)
M(t)}\right]
\nonumber \\
&=& {\rm E}\left[\frac{dM }{ M} + \frac{dp}{p} + \frac{dM}{M}
\frac{dp}{ p}\right]~. \label{mgmll;s} \ea

\subsection{Price dynamics with rational expectation (RE) and the
stochastic discount factor}

The price dynamics is written in the standard form
\be
\frac{dp}{p} = \mu(t) dt +
\sigma(t) dW - \kappa dj  ~, \label{ghgnaql}
\ee
where $dW$ is the
increment of a random walk with no drift and unit variance and
$dj$ is the jump process equal to $0$ in absence of a crash and
equal to $1$ when a crash occurs. The crash hazard rate $h(t)$ is
given by
\be
{\rm E}_t[dj]=h(t) dt
\label{mngml}
\ee
by definition, where ${\rm E}_t[x]$ denotes the expectation of $x$
conditioned on all information available at time $t$.

To obtain the RE price dynamics, we determine $\mu(t)$ such
that the the process $p M$ be a martingale, that is,
the factor proportional to $dt$ of the last term of (\ref{mgmll;s})
\be
\mu(t) - r(t) - \kappa h(t) - \sigma(t) \phi(t)~,
\ee
be zero. This leads to
\be
\mu(t) - r(t) = \kappa h(t) + \sigma(t) \phi(t)~, \label{gblanvla}
\ee
where we
have used that ${\rm E}_t [d{\hat W}]=0={\rm E}_t [dW \cdot d{\hat
W}]$, ${\rm E}_t [d{\hat W}^2] =dt$ and (\ref{mngml}).
The expected price conditioned on no crash/rally occurring
($dj=0$) is obtained by integrating (\ref{ghgnaql}) with
(\ref{gblanvla}):
\be
{\rm E}_{t_0}[p(t)] = p(t_0) ~L(t) ~ \exp
\left( \kappa \int_{t_0}^t d\tau  h(\tau) \right)~,
\label{Eq:Ep}
\ee
where
\be
L(t) = \exp \left(\int_{t_0}^t d\tau  \left[ r(\tau)
+ \sigma(\tau) \phi(\tau) \right] \right)~.
\label{Eq:L}
\ee
For $r(t)=\phi(t)=0$, $L(t)=1$, we recover the previous formulation
\cite{Johansen-Sornette-1999-Risk,Johansen-Sornette-Ledoit-1999-JR,Johansen-Ledoit-Sornette-2000-IJTAF}.

Expression (\ref{Eq:Ep}) with (\ref{Eq:L}) represents the dynamics of the asset
price $p(t)$ as the results of three contributions:
\begin{itemize}
\item the interest rate $r(t)$, which provides as usual the reference
background
against which to compare stock market price returns,

\item the price volatility $\sigma(t)$ priced
in proportion of the market price of risk $\phi(t)$ of the SDF, the later
embodying the positive relation between risk and return suggested
by most asset pricing models \cite{Merton1973},

\item the crash hazard rate, through another risk-return positive
relationship emerging from the negative price drop associated
with a crash.
\end{itemize}

\subsection{Fit with the LPPL model for the crash hazard rate $h(t)$
\label{mgmjle}}

Knowing the empirical data of the interest rate $r(t)$ and the volatility,
we use equation (\ref{Eq:Ep}) with (\ref{Eq:L}) augmented
with a model of the crash hazard rate
(and assuming a simple dynamics for $\phi(t)$) to fit
the empirical price time series $p(t)$.
In this section, we use the simplest specification that
$\phi(t)=\phi$ is constant.
Here, we model the crash hazard rate as explained in
\cite{Johansen-Sornette-Ledoit-1999-JR,Johansen-Ledoit-Sornette-2000-IJTAF}
(for a general presentation and specific derivation
of the crash hazard rate from herding, see \cite{crashbook}).
In a nutshell, the crash hazard rate is modeled by a log-periodic power
law (LPPL) which captures the intermittent positive feedback during
imitation regimes of bubbles before crashes and antibubbles developing
after a bubble. This leads to modeling
the integral of $h(t)$ in the exponential term with the following LPPL formula
\begin{equation}
\kappa\int_{t_0}^t{\rm{d}}\tau h(\tau) = B|t_c-t|^m +
C|t_c-t|^m\cos[\omega\ln|t_c-t|-\psi]~.\label{Eq:inth}
\end{equation}
Putting $\sigma(t)$ and $r(t)$ to zero
retrieves previous works consisting in fitting the logarithm of the price by
the LPPL formula.

Let us rewrite the formula (\ref{Eq:Ep})
and (\ref{Eq:L}) in the following form
\begin{equation}
y(t) \equiv \ln[p(t)]-\int_{t_0}^t r(\tau){\rm{d}}\tau =
\ln[p(t_0)] + \phi \int_{t_0}^t \sigma(\tau) {\rm{d}}\tau +
\kappa\int_{t_0}^t{\rm{d}}\tau
h(\tau) \label{Eq:logp}
\end{equation}
where we replace $E_{t_0}[p(t)]$ by the realized price $p(t)$.
Substitution of Eq.~(\ref{Eq:inth}) into (\ref{Eq:logp}) results in
\begin{equation}
y(t) = A + \phi v(t) + B f(t) + C g(t)~, \label{Eq:LPPL}
\end{equation}
where
\begin{equation}
\left\{
\begin{array}{rcl}
y(t) &=& \ln[p(t)]-\int_{t_0}^t r(\tau){\rm{d}}\tau~,\\
A    &=& \ln[p(t_0)]~,\\
v(t) &=& \int_{t_0}^t \sigma(\tau){\rm{d}}\tau~,\\
f(t) &=& |t_c-t|^m~,\\
g(t) &=& |t_c-t|^m\cos[\omega\ln|t_c-t|-\psi]~.
\end{array}\right.
\label{Eq:yAvfg}
\end{equation}

We use the following procedure.
\begin{enumerate}
\item We take $p(t)$ to represent the S\&P500 index price.

\item We specify $r(t)$ as being the yield of the three-month
Treasury bill.

\item We take the CBOE Volatility Index (VIX) as the proxy
for the volatility factor $\sigma(t)$, which is one of the world's
most popular measures of investors' expectations about future stock
market volatility (that is, risk)\footnote{See
$http://www.cboe.com/$ for more information on the VIX.}.

\item $t_0$ is the beginning time of the interval over which the
reduced price time series $y(t)$ is fitted. It is chosen a priori
as explained below.

\item We perform an OLS fit of $y(t)$ with the free parameters $A,
B, C, \phi, t_c, m, \omega, \psi$. Note that $A, B, C$ are linear
parameters which can be slaved to the others. $\phi$ is not a linear
parameter because it must remain positive or zero\footnote{It is
however possible to slave $\phi$ as a linear parameter and then
check at the end of the OLS fit if it is positive. If yes, we accept
the fit, if not, we put $\phi=0$ and then redo the fit without this
parameter.}
\end{enumerate}

We use the trapezoid scheme for integration of
$\int_{t_0}^t r(\tau){\rm{d}}\tau$ and
$\int_{t_0}^t\sigma(\tau){\rm{d}}\tau$:
\begin{equation}
\left\{
\begin{array}{rcl}
y(t) &=& \ln[p(t)]-\sum_{\tau=t_0+1}^t \left[r(\tau-1)+r(\tau)\right]/2~\\
v(t) &=& \sum_{\tau=t_0+1}^t \left[\sigma(\tau-1)+\sigma(\tau)\right]/2~
\end{array}\right.
\label{Eq:yAvfg:Molel1}
\end{equation}
Using Eqs.~(\ref{Eq:LPPL},\ref{Eq:yAvfg},\ref{Eq:yAvfg:Molel1}), we
have fitted the USA S\&P 500 price in its antibubble regime since
year 2000 (see \cite{SZ2002,ZS1,ZS2,ZS3} for previous related works
on the behavior of the market after the crash of March-April 2000)
and the result is shown in Fig.~\ref{Fig:NewLPPL:M1}. The r.m.s. of
the fit residuals is $\chi = 0.20$, which is very large compared
with the pure LPPL fit, showing that Model 1 is a bad representation
of the price.

In addition to the large value of the r.m.s. of the fit residuals, there
is another diagnostic for the bad quality of Model 1: the three
different contributions (LPPL, VIX and interest rate $r(t)$) are large
compared with their sum, suggesting the
existence of spurious (in the sense of ``over-fitting'') compensations between
these terms. In contrast, since the LPPL term was shown
to fit very well the data over the period from 2000 to 2003
\cite{SZ2002,ZS1,ZS2,ZS3},
the LPPL term should be the
leading contribution while the other factors should
have been perturbations, perhaps growing with time. In other words,
Model 1 is not a perturbation of the LPPL model used previously
in \cite{SZ2002,ZS1,ZS2,ZS3}. And the reason for this is that
the interest rate contribution $\int_{t_0}^t r(\tau){\rm{d}}\tau$
is given and its amplitude cannot be modulated within the RE formulation.
Since the risk-free interest rate has been always positive, its integral
grows with time, given a contribution with a trend opposite to the
overall negative trend of the price over the period 2000 to 2003; hence,
the need for the LPPL contribution to counter-balance this effect and
the ensuing bad fit and probably strong over-fitting.

\section{Model 2: Weaker semi-martingale condition on prices}

\begin{figure}
\centering
\includegraphics[width=14cm]{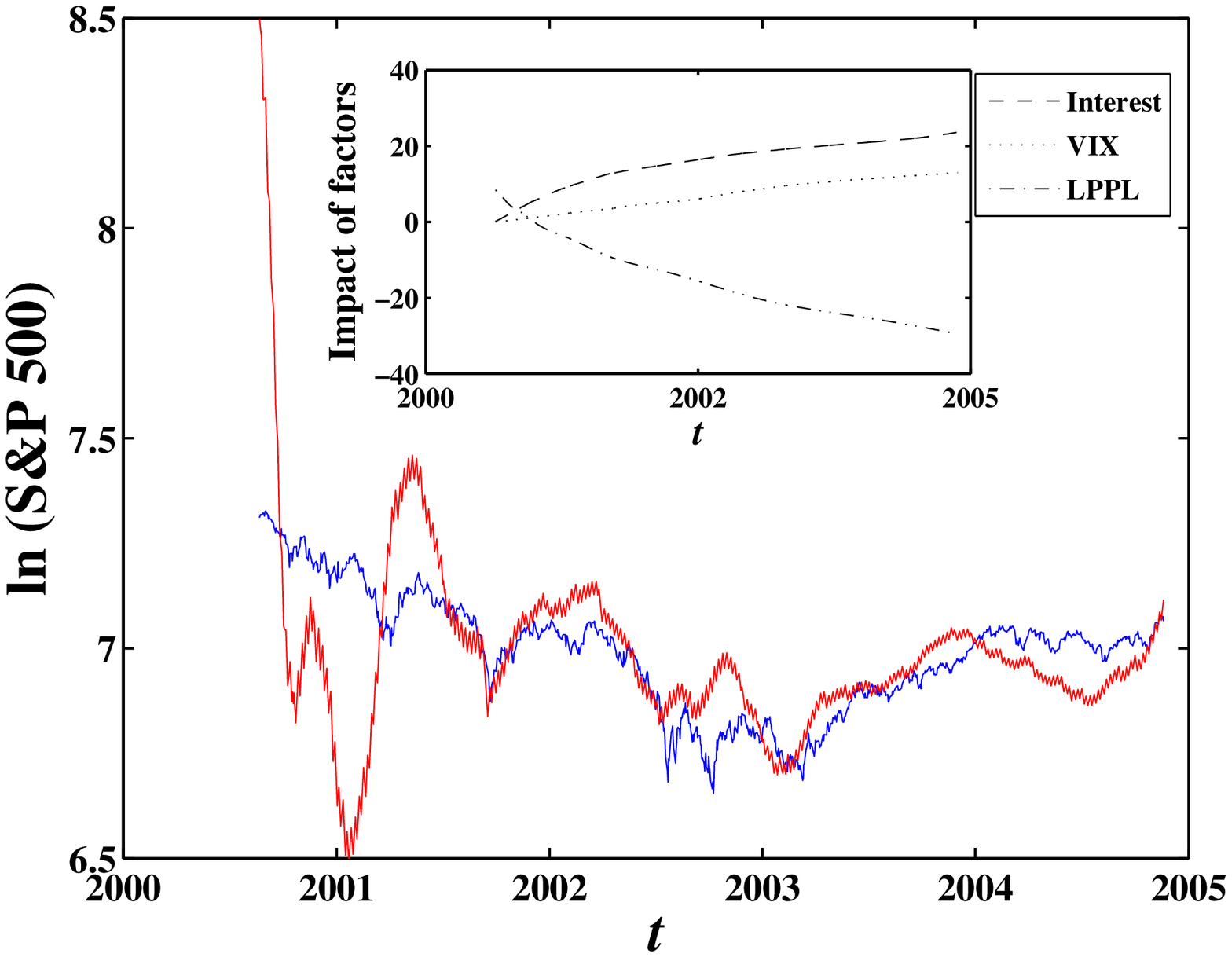}
\caption{\label{Fig:NewLPPL:M1} Comparison of the S\&P500 index
price with the best fit obtained with Eq.~(\ref{Eq:logp}),
Eq.~(\ref{Eq:LPPL}) and Eq.~(\ref{Eq:yAvfg:Molel1}). The fitted
parameter values are the following: $t_c = {\rm{2000/07/09}}$, $m =
0.44$, $\omega = 7.83$, $\psi=2.78$, $A=18.70$,
$\phi=5.16\times10^{-4}$, $B=-1.93$, and $C=-0.03$. The inset shows
the impact of different factors.}
\end{figure}

Generalizing Model 1 to include other factors does not improve,
because the positive contribution of the risk-free interest rate
(which is an essential component in the RE formulation of the
pricing model) requires the other terms and in particular the LPPL
component to adjust to large negative values to compensate its large
positive contribution. The risk-free interest rate turns out to have
an overwhelming and counter trending role in the bearish phase of
the market with declining prices. In the context of the RE
formalism, a declining price would then require very strong
antagonistic forces to compensate, leading to unrealistic impacts of
other factors and of the LPPL contribution. Adding other factors do
not provide realistic explanations of  the price trajectory over
this period. We do not show the fits and corresponding figures
obtained with all the factors we have tested, as they teach a lesson
similar to that of Fig.~\ref{Fig:NewLPPL:M1}. We conclude that it is
essentially impossible to obtain a good model of the empirical price
trajectory using the no-arbitrage rational expectation bubble model
enriched with factors such as interest rate spread, volatility and
exchange rates.

We are thus forced to remove the no-arbitrage condition and
replace it by a weaker condition. This amounts to assuming that
the market is incomplete. Indeed, we find that the technical
source of the problems in fitting the empirical data with our
model is the very large cumulative contribution of the risk-free
interest rate $r(t)$. Recall that this contribution comes from the
no-arbitrage condition obtained under a change of probability
measure from the objective to a `risk-neutral' probability measure
(a priori distinct from the objective one), which changes the
price process from a semi-martingale into a martingale
\cite{kreps,pliska}. We thus propose to change the no-arbitrage
condition, which is equivalent to requiring that the prices are
martingales, by the condition that the prices follow a
semi-martingale, with the drift of the semi-martingale taken
proportional to the risk-free rate. In the setting of incomplete
markets, the fair price is not attainable as the particular
expectation over a unique risk-neutral measure, but rather as a
supremum over an infinite set of equivalent martingale measures
\cite{KingKorf}. Our specification for the drift of our
semi-martingale price amounts to reducing the set of equivalent
measures as in a variational set-up. We allow this drift to adjust
its amplitude to reflect the existence of price deviations which
have not been arbitraged away, possibly due to technical
constraints such as transaction costs and short-sale constraints,
but also because of the reluctance of arbitragers to take risks
against the crowd sentiments. In other words, the abandon of
the no-arbitrage condition reflects the non-fully rational
behavior of investors.

\begin{figure}
\centering
\includegraphics[width=14cm]{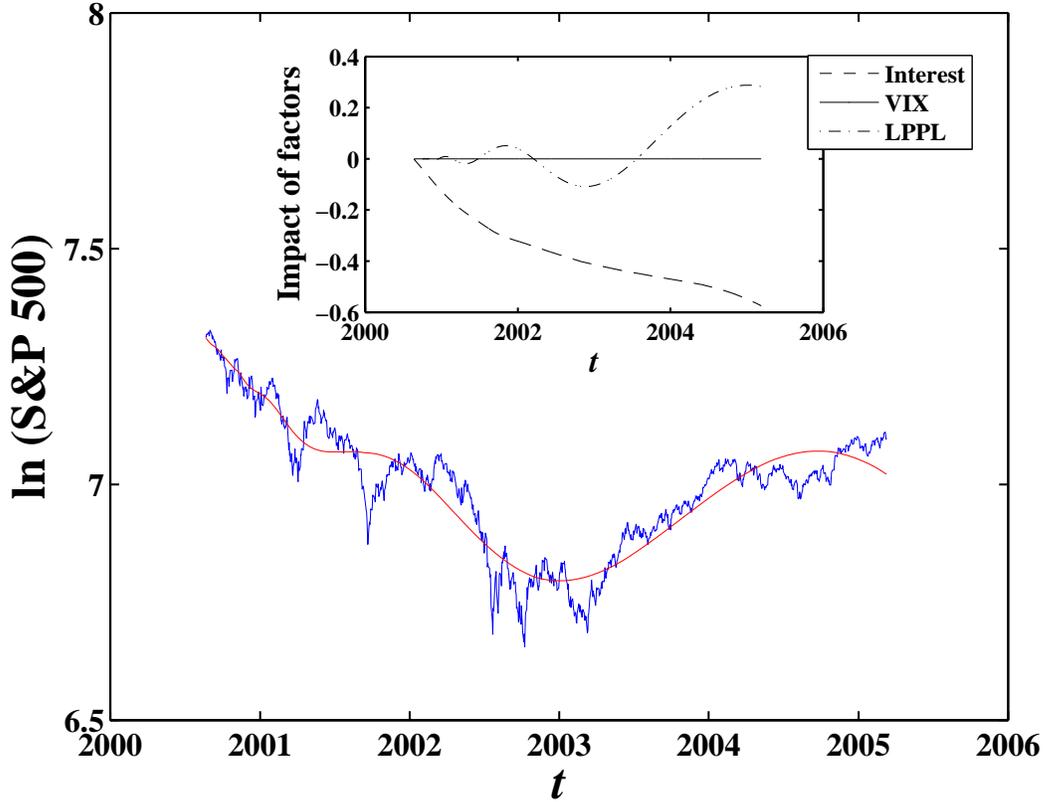}
\caption{\label{Fig:NewLPPL:M2} Comparison of the S\&P500 index
price with the best fit of model 2 defined by expression
(\ref{Eq:Ep}) with (\ref{Eq:NeoLPPL:M2}) and (\ref{Eq:inth}). The
fitted parameter values are the following: $t_c = 2000/09/30$; $m =
1.22$, $\omega = 4.68$, $\psi=3.62$, $\phi=0$, $A=7.32$,
$\alpha=-1.58\times10^{-2}$, $B=4.58\times10^{-6}$, and
$C=3.51\times10^{-5}$. The r.m.s. of the fit residuals is 0.0482.
The inset shows the impact of different factors. The the LPPL factor
is reduced or translated by the amount of $A$ for a better
comparison.}
\end{figure}

The transition to incomplete markets amounts to change (\ref{Eq:L})
into \be L(t) = \exp \left(\int_{t_0}^t d\tau  \left[ \alpha r(\tau)
+ \phi \sigma(\tau) \right] \right)~, \label{Eq:NeoLPPL:M2} \ee
where $\alpha$ is a new adjustable parameter. Expression
(\ref{Eq:Ep}) with (\ref{Eq:NeoLPPL:M2}) and (\ref{Eq:inth}) define
our Model 2, whose fit to the price time series of the the S\&P500
already shown in Fig.~\ref{Fig:NewLPPL:M1} is presented in
Fig.~\ref{Fig:NewLPPL:M2}. The r.m.s. of the fit residuals is $\chi
= 0.0482$. The improvement of model 2 compared with model 1 is
twofold and very significant. First, model 2 better captures the
evolution of the S\&P 500 index with a much smaller value of the
r.m.s. of the fit residuals. Second, the relative amplitudes of the
impact of interest and VIX are much smaller than in model 1.
However, we see that the impact of VIX is still very large compared
with the LPPL factor. This calls for further modification of the
model.

\section{Addition of other factors}

\subsection{Description of other potentially relevant factors}

The introduction of the price volatility $\sigma(t)$ in Models 1 and
2 of the previous section was justified by the assumed positive
relation between risk and return suggested by most asset pricing
models \cite{Merton1973}. There is a large empirical literature that
has tried to establish the existence of such a tradeoff between risk
and return for stock market indices. As summarized in
\cite{Goyalsanta}, the results have been inconclusive and the
relation between risk and return is often found insignificant, and
sometimes even negative. However, with better monthly variance
estimates based on past daily squared returns, the ICAPM
intertemporal relation between the conditional mean and the
conditional variance of the aggregate stock market return has
recently been found positive \cite{Ghysels}. One possible reason for
the difficulties in measuring a positive risk-return relation is
that other state variables in addition to the volatility may
influence the investment opportunities anticipated by investors
\cite{Campbell,Scruggs}. In particular, variables related to the
business cycle have been found to predict returns
\cite{Campbellbook}. They include the dividend-price ratio, the
relative Treasury bill rate, the default spread (difference between
the yield on BAA and AAA-rated corporate bonds, obtained from the
FRED database), the Fama-French factors Rm, SMB, and
HML\footnote{See the website at
\url{http://mba.tuck.dartmouth.edu/pages/faculty/ken.french/data_library.html}},
the exchange rates and so on. The excess return Rm on the market is
calculated as the value-weighted return on all NYSE, AMEX, and
NASDAQ stocks (from CRSP) minus the one-month Treasury bill rate
(from Ibbotson Associates). SMB (Small Minus Big) is the average
return on three portfolios made of assets with small capitalizations
minus the average return on three portfolios made of assets with big
capitalizations. HML (High Minus Low) is the average return on two
value portfolios minus the average return on two growth portfolios.
In addition, determinants of investors' risk aversion identified in
the asset pricing literature are economic growth prospects, measures
of equity and credit market risk, fluctuations in the exchange rate
and negative news events in other equity markets. If any of these
factors are anticipated to forecast or influence future returns,
they should be incorporated in the return equation. We thus discuss
extensions of Model 2 which incorporate different additional factors
as now explained.

\subsection{Model 3: Adding the exchange rate factor}

The fact that exchange rates (for instance euro/\$, i.e., the value
of one euro in dollars) is a relevant factor in the price equation
is suggested empirically by a live experiment to forecast the
S\&P500 index that we have performed from Oct. 2002 till Nov. 2004.
This experiment led to the suggestion that the S\&P500 index was
more predictable when expressed in euros or other major foreign
currencies than in US dollars. The rational for this observation is
multifold. First the conversion of the dollar value of the US stock
market into euros reflects more the view point of foreign investors,
which constitute a growing fraction of the population of investors,
especially at times of speculative bubbles and more generally
herding \cite{evidence}, but also due to the growing influence of
debts at times of bulging US deficits. Second, valuing in euro
provides a simple mean of eliminating (at least in part) the dollar
effect resulting from the policy of the Fed (and in particular its
liquidity inputs). Third, the relevance of the exchange rate as an
important factor to forecast returns may reflect the expectation
that a lower dollar means a more competitive US economy, with its
many spillovers on market shares, job creation, earnings,
investments and productivity gains and so on. Inversely, a weaker
dollar may actually be worrying because it favors the resumption of
inflation, higher interest rates which may slow down the economy,
potential loss of confidence of foreign investors on the
sustainability of the US deficits, varying profits for companies
trading internationally and so on. The exchange rate is also
responding to global trade deficit of the US versus the rest of the
world. It is often advanced that the progressive collapse of the US
dollar since 2002 is in large part driven by the growing US deficits
(Federal debt, private debt, and current account)
\cite{Obstfeld-Rogoff-2004-NBER}. These effects are in general going
in opposite directions but with different time scales.

Let us thus write \be p_{\$}(t) = p_{\rm euro}(t) e^{R_{\rm
exch}(t)}~, \label{gvmkme} \ee which defines the exchange rate
return between the dollar and the euro. Putting for instance $p_{\rm
euro}=1$ gives the number of dollars equal to $e^{R_{\rm exch}}$ we
obtain from one euro at time $t$. Now, the risk perceived by the
economic agents is not determined by $e^{R_{\rm exch}(t)}$ or
$R_{\rm exch}(t)$ but by the relative temporal variations of
$p_{\$}(t) / p_{\rm euro}(t)$. We have $R_{\rm exch}(t) = \ln
p_{\$}(t) - \ln p_{\rm euro}(t)$ and thus \be r_{\rm exch}(t) \equiv
dR_{\rm exch}(t)/dt = \frac{dp_{\$}(t)/dt}{p_{\$}(t)} -
\frac{dp_{\rm euro}(t)/dt}{p_{\rm euro}(t)}~. \label{mgmklds} \ee
$r_{\rm exch}(t)$ defined in (\ref{mgmklds}) quantifies the real
risks due to changes in exchange rates. Indeed, an investor long in
dollars (say, he owns one unit of dollars) has the equivalent wealth
of $e^{-R_{\rm exch}(t)}$ euros at time $t$, which becomes
$e^{-R_{\rm exch}(t+dt)}$ euros at time $t+dt$. He makes a relative
profit or loss (return) in euro equal to \be \frac{e^{-R_{\rm
exch}(t+dt)} - e^{-R_{\rm exch}(t)}}{e^{-R_{\rm exch}(t)}} = -
r_{\rm exch}(t) dt~. \ee Of course, the situation is symmetric for
an investor having one euro and measuring his wealth variation in
dollars, leading to a profit or loss of $+r_{\rm exch}(t) dt$ over
one elementary time step $dt$.

We thus generalize the market price of risks
\begin{equation}
\phi(t) dW(t) \to \phi dW +
\eta_{\rm exch} r_{\rm exch}(t) dW_{\rm exch}~, \label{mgmellaas}
\end{equation}
where $\eta_{\rm exch}$ sets the scale of the impact of the risks
associated with the exchange rate fluctuations. Note that
$\eta_{\rm exch}$ can be either positive or negative.
Correlatively, we generalize the price dynamics (\ref{ghgnaql}) to
make explicit its dependence on the factors identified in the
market price of risks:
\begin{equation}
\frac{dp}{p} = \mu(t) dt + \sigma_p^{\rm exch} dW_{\rm exch} +
\sigma(t) dW - \kappa dj ~. \label{ghgnaqsdaaasd}
\end{equation}
The total volatility (variance) of the price returns
$[\sigma_p^{\rm exch}]^2 + [\sigma(t)]^2$ is the sum of two
contributions, one resulting from the fluctuations associated with
the exchange rate and the other which captures all other factors
of uncertainty.
Then, the semi-martingale condition leads to change
(\ref{gblanvla}) into
\begin{equation}
\mu(t) = \alpha r(t) +\kappa h(t) + \gamma \sigma(t)
+\alpha_{\rm{exch}}r_{\rm exch}(t)~, \label{gblanvladfsaasd}
\end{equation}
where $\alpha_{\rm{exch}}=\eta_{\rm exch} \sigma_p^{\rm exch}$. The
expected price conditioned on no crash/rally occurring ($dj=0$) is
of the form (\ref{Eq:Ep}) with
\begin{equation}
\L(t) = \exp \left(\int_{t_0}^t d\tau \left[
\alpha r(\tau) + \gamma \sigma + \alpha_{{\rm{exch}}} r_{\rm{exch}} \right]
\right)~. \label{Eq:L223}
\end{equation}
Note that $\gamma \geq 0$, while $\alpha_{\rm exch}$ can be of any
sign, depending whether the risk factor is $r_{\rm exch}$ or its
negative $-r_{\rm exch}$. The choice of one or the other amounts
to analyzing the exchange rate from dollars to euros or
vice-versa. This symmetry allows for an a priori arbitrary sign
for $\eta_{\rm exch}$.

According to (\ref{gvmkme}) and (\ref{mgmklds}), \be \int_{t_0}^t
d\tau ~ r_{\rm exch}(t) = \ln \left[ \frac{p_{\$}(t)}{ p_{\rm
euro}(t)} \frac{p_{\rm euro}(t_0) }{p_{\$}(t_0)} \right]~. \ee Then,
for the special choice $\alpha_{\rm{exch}}=\eta_{\rm exch}
\sigma_p^{\rm exch}=1$, we have \be e^{\alpha_{\rm{exch}}
\int_{t_0}^t d\tau ~ r_{\rm exch}(\tau)} = \frac{p_{\$}(t)}{ p_{\rm
euro}(t)} \frac{p_{\rm euro}(t_0)}{p_{\$}(t_0)}~. \ee In this case,
expression (\ref{Eq:Ep}) with $p(t)$ understood as $p_{\$}(t)$ is
replaced by \be {\rm E}_{t_0}[p_{\rm euro}(t)] = p_{\rm euro}(t_0)
~{\hat L}(t) ~ \exp \left( \kappa \int_{t_0}^t d\tau  h(\tau)
\right)~, \label{Eq:Epeuro} \ee where \be {\hat L}(t) = \exp
\left(\int_{t_0}^t d\tau  \left[\alpha r(\tau) + \gamma \sigma(\tau)
\right] \right)~. \label{Eq:L22afasa} \ee In words, equation
(\ref{Eq:Epeuro}) means that the term $\exp \left(\int_{t_0}^t d\tau
\left[ \alpha_{{\rm{exch}}} r_{\rm{exch}} \right]\right)$ in
(\ref{Eq:L223}) embodying the exchange risk factor is absorbed in
the transformation from dollars to euros. Model 3 thus provides a
justification for expressing the US stock market from the view point
of the euro as we have done in our monthly online predictions (our
monthly predictions used ${\hat L}(t)=1$). This procedure is nothing
but the particular case ($\alpha_{\rm{exch}}=\eta_{\rm exch}
\sigma_p^{\rm exch}=1$) of Model 3 which incorporates the risk
factor associated with the exchange rates between the US dollar and
other currencies. By generalizing the description of the US stock
market viewed not solely from the vantage of foreigners, Model 3
allows for a market price of risk associated with the exchange rate
which is not entirely described in terms of the euro view point.

We implement Model 3 numerically as:
\begin{equation}
y(t) = A + \alpha r(t) + \gamma v(t)+
\alpha_{\rm exch} R(t) + B f(t) + C g(t)~, \label{Eq:LPPL:model:2}
\end{equation}
where
\begin{equation}
\left\{
\begin{array}{rcl}
y(t) &=& \ln[p(t)]~,\\
A    &=& \ln[p(t_0)]~,\\
r(t) &=& \int_{t_0}^t r_{3m}(\tau){\rm{d}}\tau~,\\
v(t) &=& \int_{t_1}^t \sigma(\tau){\rm{d}}\tau~,\\
R(t) &=& \int_{t_0}^t r_{\rm{exch}}(\tau){\rm{d}}\tau~,\\
f(t) &=& (t_c-t)^m~,\\
g(t) &=& (t_c-t)^m\cos[\omega\ln(t_c-t)-\phi]~,
\end{array}\right.
\label{Eq:yAvfg:model:2}
\end{equation}

\begin{figure}
\centering
\includegraphics[width=14cm]{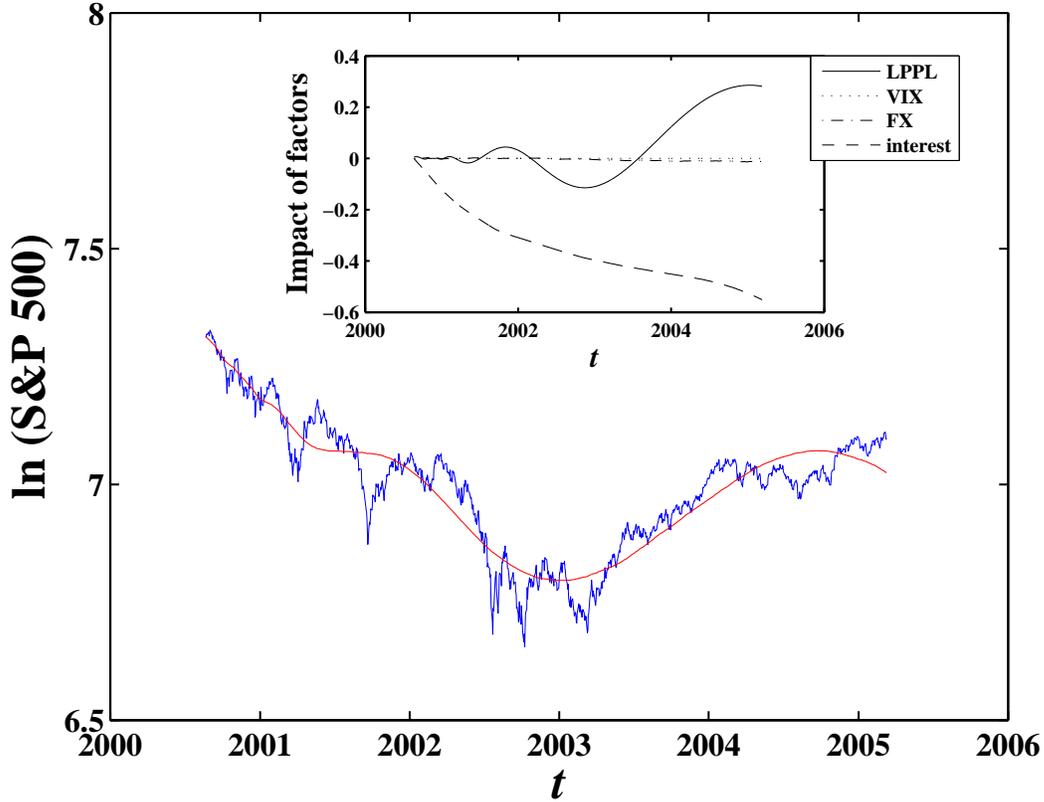}
\caption{\label{Fig:NewLPPL:M3} Comparison of the S\&P500 index
price with the best fit of model 3 defined by expression
(\ref{Eq:Ep}) with (\ref{Eq:NeoLPPL:M2}) and (\ref{Eq:inth}). The
fitted parameter values are the following: $t_c = 2000/10/17$; $m =
1.25$, $\omega = 4.52$, $\psi=4.83$, $\gamma=0$, $A=7.31$,
$\alpha=-1.55\times10^{-2}$, $\alpha_{\rm{exch}}=2.02\times10^{-2}$,
$B=2.68\times10^{-6}$, and $C=2.98\times10^{-5}$. The r.m.s. of the
fit residuals is 0.0482. The inset shows the impact of different
factors. The LPPL factor is reduced or translated by the amount of
$A$ for a better comparison.}
\end{figure}

Figure \ref{Fig:NewLPPL:M3} shows the evolution of the S\&P 500
index from Aug-20-2000 to Mar-21-2005 fitted with Model 3. The
r.m.s. of the fit residuals is 0.0482. We notice that this figure is
undistinguishable from Fig.~\ref{Fig:NewLPPL:M2}. Both model 2 and
model 3 gives the same r.m.s. of fit residuals and the fitted
parameters are quite close to each other. In the inset, we see that
the impact of the foreign exchange rate can be neglected.
Notwithstanding the decrease of the interest rate during the
antibubble period, which is causally slaved by the market
\cite{ZS2004}, the short term interest rate has increasing negative
impact on the stock market.

\begin{figure}
\centering
\includegraphics[width=14cm]{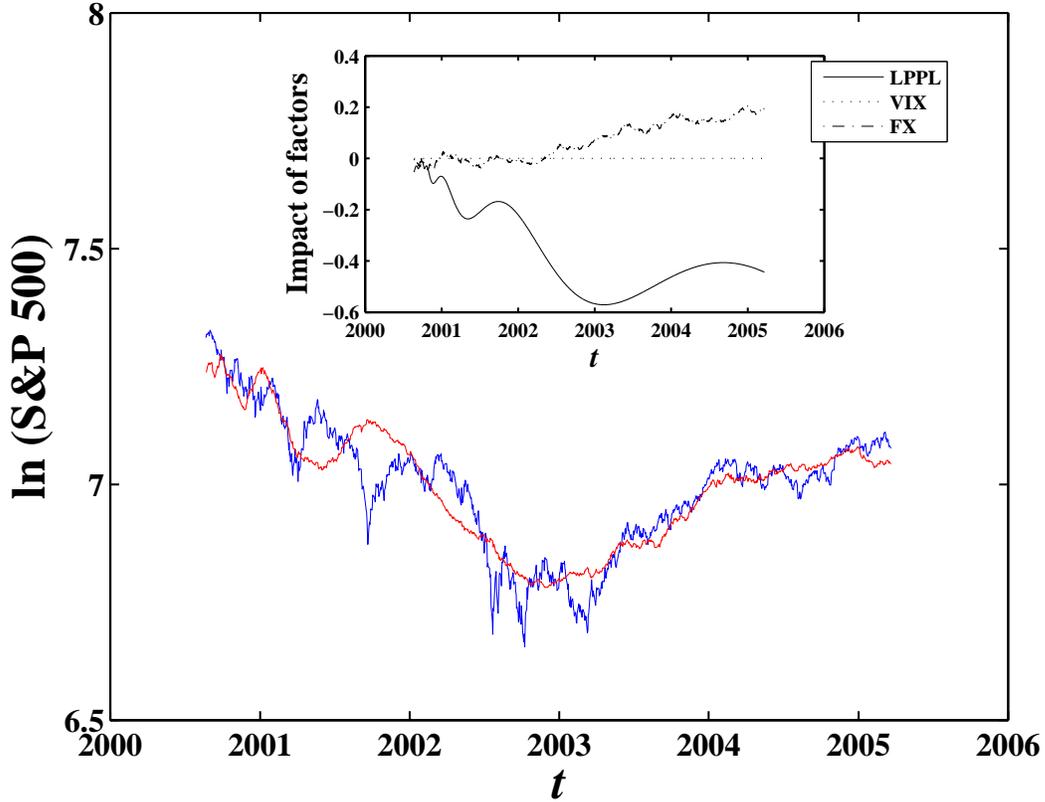}
\caption{\label{Fig:NewLPPL:M3b} Comparison of the S\&P500 index
price with the best fit of the modified model 3. The fitted
parameter values are the following: $t_c = 2000/09/25$; $m = 0.65$,
$\omega = 4.60$, $\psi=3.99$, $\gamma=0$, $A=7.29$,
$\alpha_{\rm{exch}}=-0.502$, $B=-5.55\times10^{-3}$, and
$C=1.93\times10^{-3}$. The r.m.s. of the fit residuals is 0.0553.
The inset shows the impact of different factors. The LPPL factor is
reduced or translated by the amount of $A$ for a better comparison.}
\end{figure}

  From Fig.~\ref{Fig:NewLPPL:M3}, the impact of interest rate is still
very large, we thus fit a modified model 3 containing LPPL, VIX, FX
but without the interest rate $r(t)$ (forcing $\alpha=0$). The
result is shown in Fig.~\ref{Fig:NewLPPL:M3b}. This new figure
emphasizes just the effect of the FX, without the interfering effect
of the interest rate $r(t)$ and again the impact of VIX can be neglected.

\subsection{Model 4: Semi-Martingale bubble model with additional factors}

It is natural to extend Model 2 defined by (\ref{Eq:Ep}) with (\ref{Eq:L223}),
by adding other known factors of risks. We explore the relevance of
the interest rate spread and of the historical volatility.
This amounts to change (\ref{Eq:L223}) into
\begin{equation}
\L = \exp \left(\int_{t_0}^t d\tau \left[\alpha r +  \gamma \sigma
+ \alpha_{\rm exch} r_{\rm exch} + \alpha_{\rm hist} \sigma_{\rm
hist} + \alpha_{\rm spread} (r_{\rm{10y}} - r_{\rm{3m}}) \right]
\right) \label{Eq:5factor}
\end{equation}
where $\alpha_{\rm hist}$ and $\alpha_{\rm spread}$ are two new
adjustable parameters. Note again that $\gamma$ and $\alpha_{\rm
hist}$ are taken positive. $r_{\rm{10y}} - r_{\rm{3m}}$ is the
difference in yields between the Treasury bill with $10$ year and
$3$ month maturities.

\begin{figure}
\centering
\includegraphics[width=14cm]{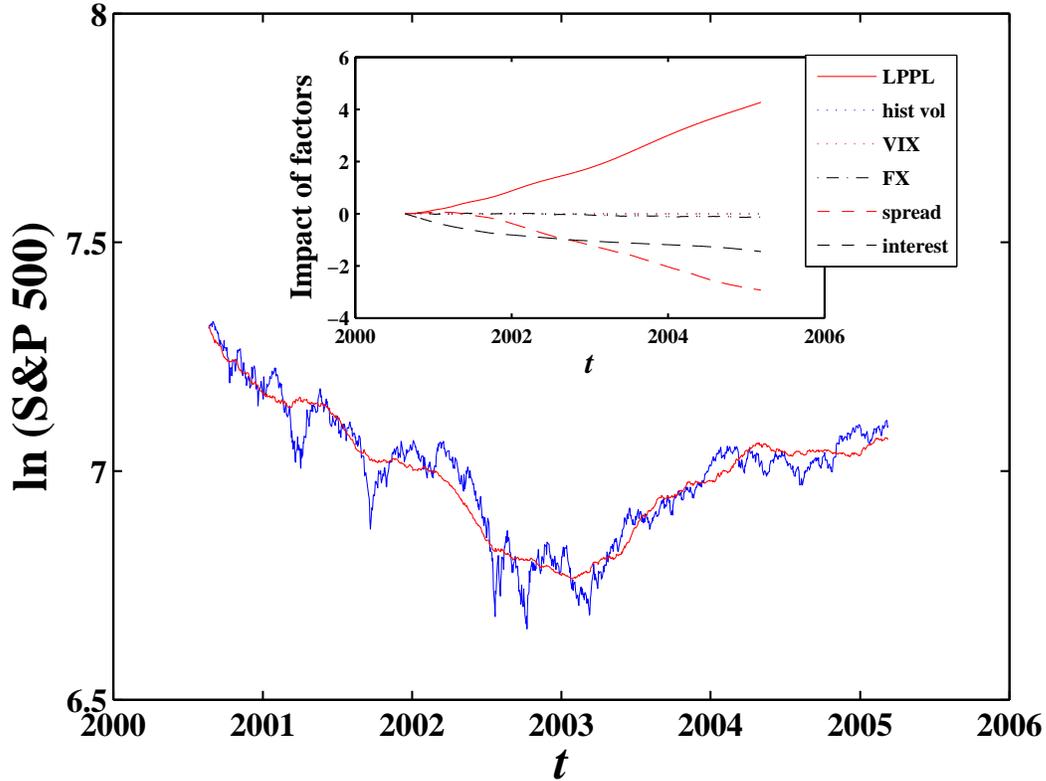}
\caption{\label{Fig:NewLPPL:M4} Evolution of the S\&P 500 index from
Aug-20-2000 to Nov-20-2003 with the fit to Model ``weak''. The
values of the fit parameters are the following: $t_c = 2000/09/09$;
$m = 1.30$, $\omega = 7.80$, $\psi=0.82$, $\gamma=0$,
$\alpha_{\rm{hist}}=0$, $A=7.31$, $\alpha=-3.89\times10^{-2}$,
$\alpha_{\rm{spread}}=-7.64\times10^{-2}$,
$\alpha_{\rm{exch}}=0.345$, $B=2.83\times10^{-4}$,
$C=9.88\times10^{-6}$. The r.m.s. of the fit residuals is 0.0409.
The inset shows the impact of different factors contributing to the
overall fit of Model 4. The LPPL factor is reduced or translated by
the amount of $A$ for a better comparison.}
\end{figure}

Proceeding as before for the OLS fit of this Model 4 to the S\&P500
price trajectory, we obtain Fig.~\ref{Fig:NewLPPL:M4}. The r.m.s. of
the fit residual is $0.0409$, which is smaller than the value
$0.0436$ of the second-order Landau fit to the same data set (which
was up to now the best, i.e., smallest r.m.s) and of course much
smaller than the r.m.s. $0.0601$ of fit residuals using the
first-order LPPL model. The inset shows the amplitudes of the
contributions of the different factors to the global fit. It shows
that, in addition to the LPPL factor previously considered in
\cite{SZ2002,ZS1,ZS2,ZS3}, the short-term interest rate and the
interest rate spread are the dominant factors, while the other three
factors are weaker. The contribution of $\sigma$ (or VIX) and the
historical volatility are actually completely negligible since the
values of $\gamma$ and $\alpha_{\rm{hist}}$ is very small and close
to zero.

Notwithstanding the overall quality of this fit shown in
Fig.~\ref{Fig:NewLPPL:M4},
we argue that it should be disqualified for the same reasons
as discussed in section \ref{mgmjle} concerning Fig.~\ref{Fig:NewLPPL:M1}:
the spread and interest rate contributions are quite large, which forces the
LPPL contribution to {\it increase} with time. This LPPL contribution
thus exhibits a trend opposite to the initial simple LPPL fit and,
therefore, Model 4
cannot be considered as a perturbation to the simple LPPL model.

\subsection{Second-order Landau formula}

An extension of the LPPL formulas (\ref{Eq:inth}) and (\ref{Eq:logp})
described in section \ref{mgmjle} has been proposed in
\cite{SJ97PhysA}, based on a formulation of the critical
power law behavior in terms of a Landau-like expansion
\begin{equation}
   \frac{dF(\tau)}{d\ln \tau} = \alpha F(\tau)+\beta |F(x\tau)|^2F(\tau)...~,
   \label{Eq:Landau0}
\end{equation}
where the coefficients are generally complex and $\tau=|t-t_c$. Keeping only
the first term $\alpha F(\tau)$ of the right-hand-side of (\ref{Eq:Landau0})
with $\alpha = m + i\omega$ retrieves (\ref{Eq:LPPL}) without
the term $\phi v(t)$, which is the so-called first-order
LPPL formula \cite{crashbook}. Including the second-order term
$\beta |F(\tau)|^2F(\tau)$ leads to \cite{SJ97PhysA}
\begin{equation}
I(t) = A + \frac{B\tau^m+ C\tau^m\cos\left\{\omega\ln \tau +
\frac{\Delta_\omega}{2m}\ln \left[1+\left(\frac{\tau}
{\Delta_t}\right)^{2m}\right] +\phi\right\}} {\sqrt{1+
\left(\frac{\tau} {\Delta_t}\right) ^{2m}}}~, \label{Eq:Landau2}
\end{equation}
which was used to model the bubbles before the USA 1929 crash and 1987
crash \cite{SJ97PhysA}. A third order LPPL formula was
used to model the nine-year Nikkei antibubble since January 1990
and predict the ensuing two-year evolution \cite{JS99IJMPC,JS00mfk}.

Equation (\ref{Eq:Landau2}) predicts the transition from the angular
log-frequency $\omega + \Delta_\omega$ for $\Delta_t < |t-t_c|$  to the
angular log-frequency $\omega$ close to $t_c$. This corresponds to an
{\it approximate} description of a log-frequency modulation.
For instance, the 1990 Nikkei antibubble studied in \cite{JS99IJMPC,JS00mfk}
experienced the transition from the first-order Landau description
(\ref{Eq:inth}) and (\ref{Eq:logp}) to the
second-order Landau formula (\ref{Eq:Landau2})
approximately 2.5 years after the inception of the antibubble
\cite{JS99IJMPC,JS00mfk}.
In contrast, the 2000 S\&P 500 antibubble was found to
just barely enter the second-order Landau
regime on the fourth quarter of 2003 after more three years since its
inception \cite{ZS3},
while the second-order Landau regime was undetectable from data ending in the
last quarter of 2002 \cite{ZS1}.

In the following tests, we compare between them the different models
constructed with
different factors and with the second-order LPPL formula (\ref{Eq:Landau2}).

\section{Tests on the relevance of the different factors}

Let us consider the family of one-factor LPPL models defined by
\begin{equation}
\L = \exp \left(\int_{t_0}^t \alpha_i f_i  d\tau\right),
~{\rm{with}}~i=1, 2, 3, 4, {\rm{and}}~5, \label{Eq:1factor}
\end{equation}
where $\alpha_1f_1=\alpha r$, $\alpha_2f_2=\gamma \sigma$,
$\alpha_3f_3=\alpha_{\rm exch} r_{\rm exch}$,
$\alpha_4f_4=\alpha_{\rm hist} \sigma_{\rm hist}$, and
$\alpha_5f_5=\alpha_{\rm spread} (r_{\rm{10y}} - r_{\rm{3m}})$. We
fit the S\&P 500 antibubble after the burst of the ``new economy'' bubble in
April of 2000 using each of the five models defined by
Eq.(\ref{Eq:1factor}). The time series is fitted from Aug-20-2000 to
a time $t_{\rm{last}}$ which is varied from
08-Apr-2002 to 10-Mar-2005 with a step of 21 trading days. Each interval
from Aug-20-2000 to a time $t_{\rm{last}}$ is fitted by each of the five
models.

\subsection{Akaike's AIC criterion}

To compare these five models with the first-order LPPL and second-order
LPPL, we use
Akaike's minimum AIC estimation \cite{Akaike}, which was designed for the
identification of several competing models. The AIC is defined by
\begin{equation}
{\rm{AIC}} = -2\ln(L)+2\kappa~, \label{Eq:AIC}
\end{equation}
where $L$ is the maximum likelihood of a given model
given the data and $\kappa$ is the number of
independently adjusted parameters within a model. By assuming a
Gaussian distribution of the fit residuals, we have
\begin{equation}
\ln L = -n\ln(\hat{\sigma})-\frac{n}{2}[\ln(2\pi)+1]~, \label{Eq:Likelifa}
\end{equation}
where $n$ is the number of data points and $\hat{\sigma}^2=\chi^2$
is the maximum likelihood estimate of the variance of the fit
residual.

\begin{figure}[h]
\centering
\includegraphics[width=14cm]{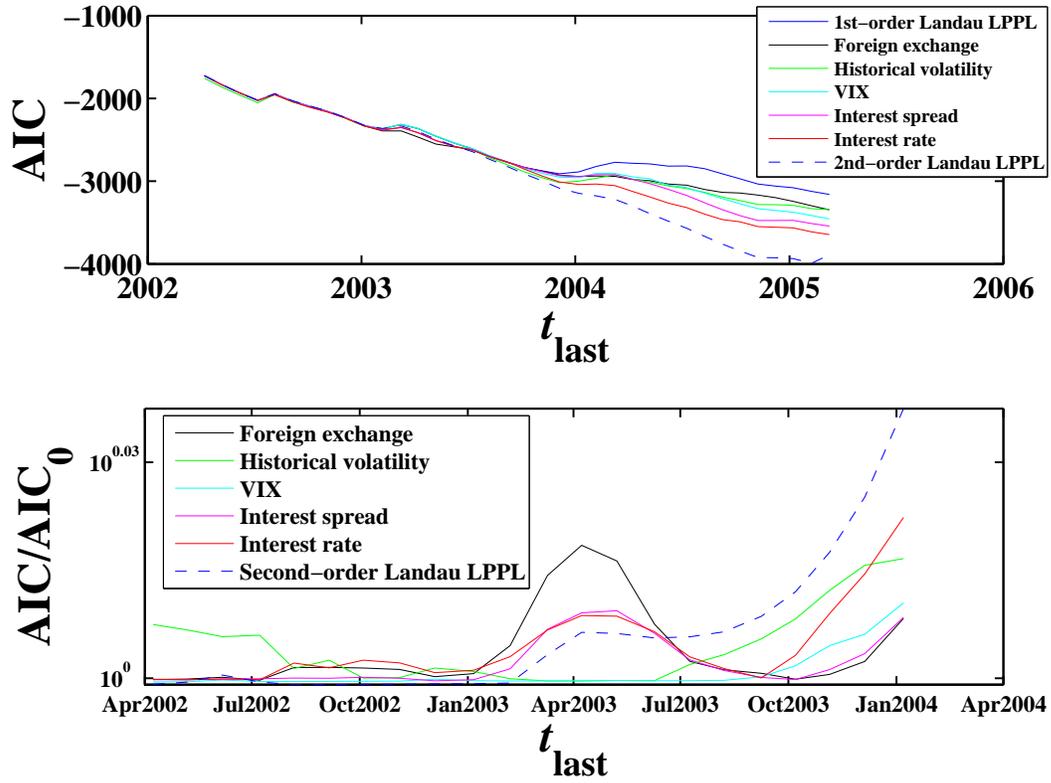}
\caption{\label{Fig:NeoLPPL:1f:AIC} (Top panel) AIC of the five
one-factor LPPL models with comparison to the first- and
second-order Landau LPPL models. (Bottom panel) Relative AIC
against the AIC$_0$ of the first-order LPPL model.}
\end{figure}

The top panel of
Fig.~\ref{Fig:NeoLPPL:1f:AIC} shows the evolution of the AIC's of the
five one-factor LPPL models in comparison with the first- and
second-order Landau LPPL models. To better illustrate the
competition among the models, we plot the relative AIC with
respect to the first-order LPPL model in the lower panel of
Fig.~\ref{Fig:NeoLPPL:1f:AIC}. This figure suggests
that the historical volatility played
the key role before August of 2002. Around October 2002, the
interest rate dominated. In the first six months of 2003, the
foreign exchange rate became the key factor. Since the end of 2003,
all factors have played an increasingly large role. Note also that
after July 2003,
the second-order Landau LPPL model has the largest AIC and is
identified as the best model, suggesting as discussed elsewhere
\cite{SZ2002,ZS3}
a progressive maturation of the herding effect in this ``antibubble,''
similarly to the case of Japan \cite{JS99as,JS00mfk}.

We should stress however that this qualitative similarity between
the US antibubble since 2000 and the Japanese antibubble since 1990
does not extend to the quantitative level. In \cite{ZS1} written in
2003, we stated that we could not find neither quantitative nor
qualitative differences between the US and the JP antibubbles.
Things have changed in the last two years. Our present analysis
shows that the second-order LPPL regimes of the two antibubbles are
quite different; for instance, the important parameter $\Delta
\omega$ have different signs and significantly different values; the
$\Delta_t$ 's are also very different. We conjecture that these
quantitative differences may signal deep divergences in the
mechanisms and evolutions of the two antibubbles in the present
regime described by the second-order LPPL model and beyond.

\subsection{Wilks' test of embedded hypotheses}

Since the single factor models defined with expression
(\ref{Eq:1factor}) contain Model 1 defined by (\ref{Eq:LPPL}) as a
special case $\alpha_i=0$, we can use Wilks theorem \cite{Rao} and
the statistical methodology of nested hypotheses to assess whether
the hypothesis that $\alpha_i=0$ can be rejected. By assuming a
Gaussian distribution of observation errors (residuals) at each data
point, the maximum likelihood estimation of the parameters amounts
exactly to the minimization of the sum of the square over all data
points (of number $n$) of the differences $\delta_j(i)$ between the
mathematical formula and the data \cite{Press}. According to Wilks
theorem of nested hypotheses, the log-likelihood-ratio \be T = -2
(L_0-L_1) = 2n(\ln\sigma_1 - \ln\sigma_0)~, \label{Eq:T} \ee is a
chi-square variable with $k$ degrees of freedom, where $k$ is the
number of restricted parameters \cite{Holden}. In the present case,
we have $k=1$. The Wilks test thus amounts to calculate the
probability that the obtained value of $T$ can be over-passed by
chance alone. If this probability is small, this means that chance
is not a convincing explanation for the large value of $T$ which
becomes meaningful. This implies a rejection of the hypothesis that
$\alpha_i=0$ is sufficient to explain the data and favor the fit
with $\alpha_i \neq 0$ as statistically significant.

\begin{figure}[h]
\centering
\includegraphics[width=14cm]{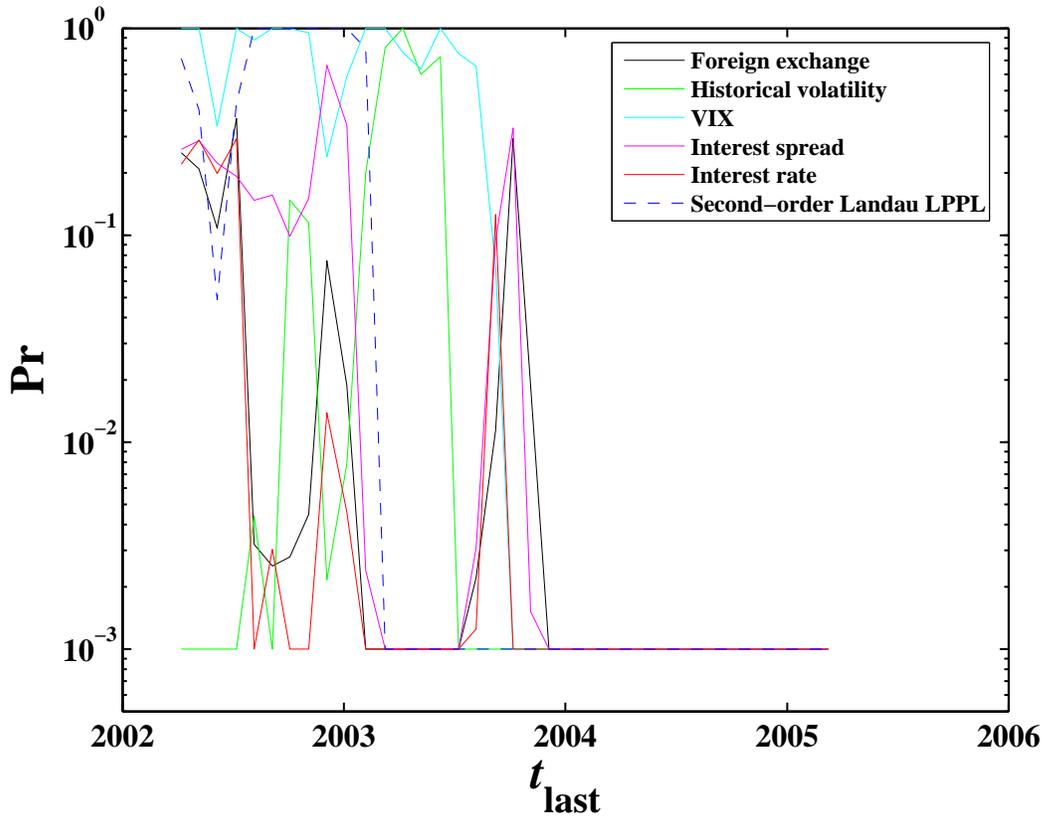}
\caption{\label{Fig:NeoLPPL:1f:Pr} Wilks' log-likelihood-ratio test
of the significance of the new factors introduced in
(\ref{Eq:1factor}). The second-order Landau formula is also tested.
Note that all values of $\rm{Pr}$ which are less than 0.001 are
plotted to be 0.001 in the figure for better representation.}
\end{figure}

Figure \ref{Fig:NeoLPPL:1f:Pr} plots the evolution of the
probability ${\rm{Pr}}$ that the log-likelihood-ratio exceeds $T$ at
$t_{\rm{last}}$. The second-order Landau formula is also tested for
comparison. We see that there are always factors that are
significant at a level of significance far better than 99\%, that
is, ${\rm{Pr}}<1\%$. For instance, $\alpha_4 \ne 0$ (historical
volatility) passed the test in the time period before August of
2002. Since 2004, all five one-factor models and the second-order
Landau model are significant.

It is noteworthy that the second-order Landau formula experienced a
transition from large ${\rm{Pr}}$ to small ${\rm{Pr}}$ in the first
quarter of 2003, signaling the crossover from the first-order Landau
regime to the second-order regime, which confirms our earlier
preliminary assessment \cite{ZS3}. In contrast, the one-factor
models exhibit large values of ${\rm{Pr}}$ for several months in
2003.

Together with the fact that the r.m.s. of the residues of the fit is
the smallest for the second-order LPPL formula (\ref{Eq:Landau2}),
we conclude that the second-order LPPL model is the best model. The
economic factors investigated here seem less powerful than just
herding effects to account for the continuation of the unraveling of
the US stock market antibubble.

\section{Prediction using the second-order LPPL formula (\ref{Eq:Landau2})}

So far, we have shown that the macroscopic factor models are not as
good as the second-order formula. For the US S\&P 500 antibubble
started in August 2000, the second-order formula (\ref{Eq:Landau2})
provides by far the best model so that
the other factors cannot explain what is going on in the US stock
market. Our analysis also
shows that the antibubble has entered the second-order Landau regime
approximately during the summer of 2003, confirming our preliminary analysis
\cite{ZS2004}. In addition, the tests in \cite{ZS2004} showed that,
in the framework of the first-order Landau model, the antibubble was
probably still alive in August 2003 but has ended since in the USA
(i.e., when viewed from the view point of a US investor valuing in
US dollars). The timing of the end of the first-order LPPL
antibubble is roughly consistent with our current analysis.

\begin{figure}[h]
\centering
\includegraphics[width=14cm]{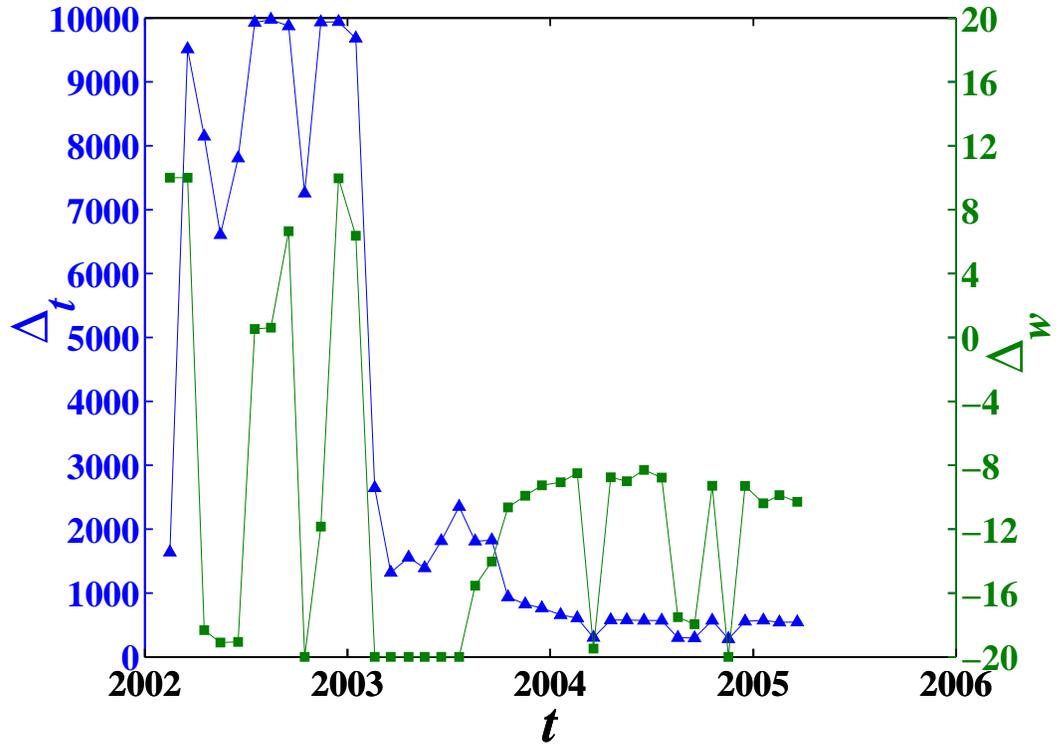}
\caption{\label{Fig:NewLPPL:DtDw} The evolution of the fitted values
of $\Delta_t$ and $\Delta_\omega$. The dramatic drop of the value of
$\Delta_t$ endorses the crossover from the first-order regime to the
second-order.}
\end{figure}

To further show that the crossover happened in 2003, we show in
Fig.~\ref{Fig:NewLPPL:DtDw} the evolution of the fitted values of
$\Delta_t$ and $\Delta_\omega$ which are the diagnostic
of the relevance of the second-order formula. The dramatic drop of the value of
$\Delta_t$ endorses the crossover from the first-order regime to the
second-order.

\begin{figure}[h]
\centering
\includegraphics[width=14cm]{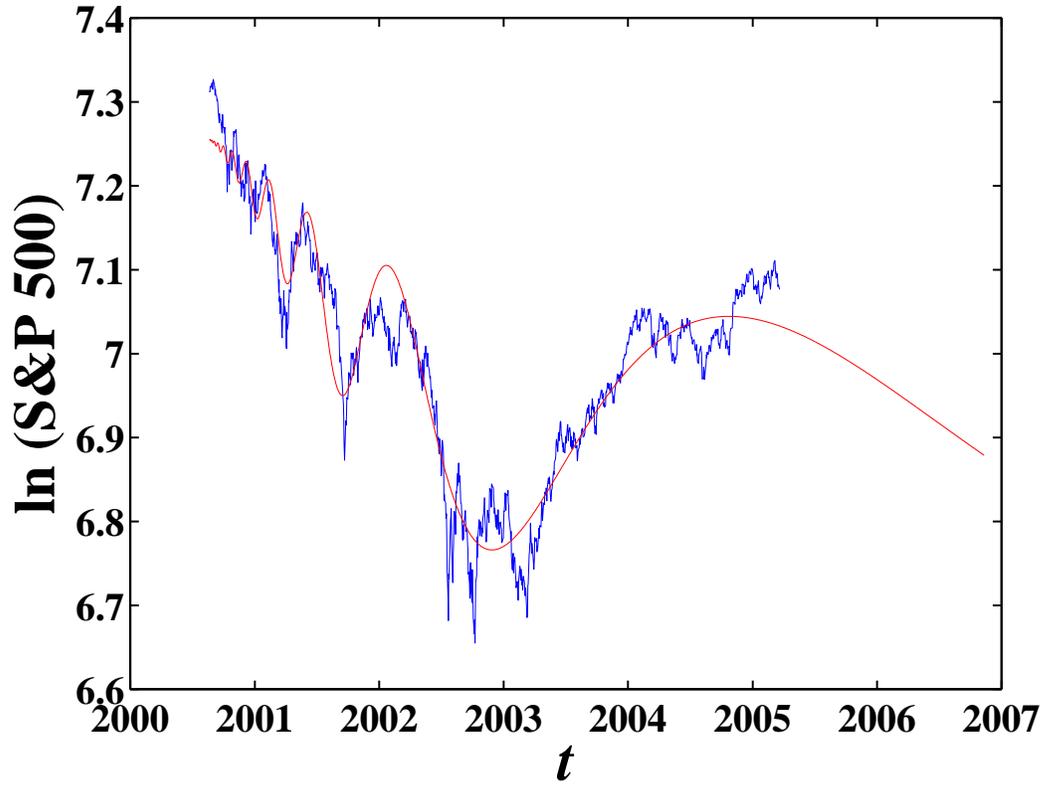}
\caption{\label{Fig:NewLPPL:Landau2Pred} Modeling and prediction of
the US S\&P 500 index from $2000/08/21$ to $2005/03/21$ using the
second-order Landau formula (\ref{Eq:Landau2}). The fitted parameters
are $t_c =
2000/08/17$, $m = 1.32$, $\omega = 13.72$, $\psi=1.31$, $\Delta_t=
545$, $\Delta_\omega=-10.28$, $A=7.26$, $B=-9.505\times10^{-5}$, and
$C=-4.224\times10^{-5}$ with the r.m.s. of fit residuals being
0.0403.}
\end{figure}

These tests suggest that it is possible to resume the modeling and
prediction of the antibubble within the second-order Landau model.
Fig.~\ref{Fig:NewLPPL:Landau2Pred} shows the fit
of the US S\&P 500 index from $2000/08/21$ to
$2005/03/21$ using the second-order Landau formula. The fitted
inception date of the antibubble $t_c = 2000/08/17$ is found stable
in all our fits with different terminal values of the fitted windows.
The extrapolation of the fitted second-order formula shown in
Fig.~\ref{Fig:NewLPPL:Landau2Pred} suggests that the market is not
yet ready for a solid recovery.

%\bibliography{NeoLPPL}

\clearpage

\clearpage

\clearpage

\clearpage

\clearpage

\clearpage

\begin{figure}
\centering
\includegraphics[width=14cm]{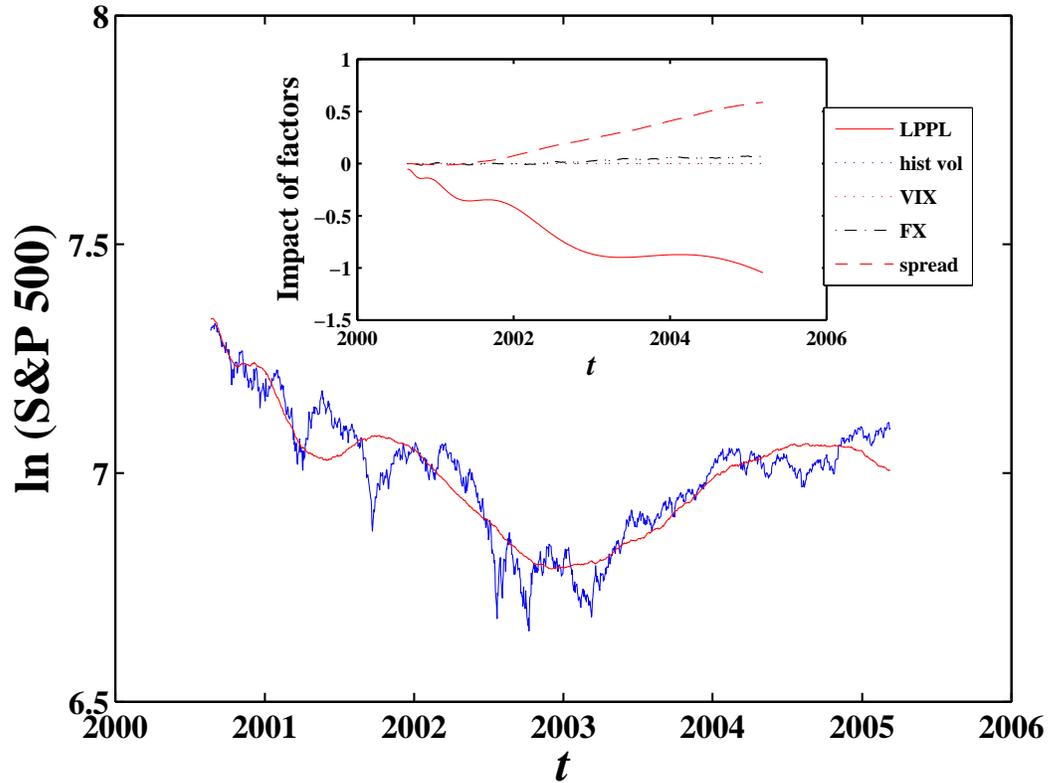}
\caption{\label{Fig:NewLPPL:M4b} Fit to the modified model of all
factors except the interest rate. The values of the fit parameters
are the following: $t_c = 2000/07/16$; $m = 0.80$, $\omega = 5.44$,
$\psi=4.17$, $\gamma=0$, $\alpha_{\rm{hist}}=9.11\times10^{-7}$,
$A=7.39$, $\alpha_{\rm{spread}}=1.54\times10^{-2}$,
$\alpha_{\rm{exch}}=-0.182$, $B=-3.23\times10^{-3}$, and
$C=5.84\times10^{-4}$. The r.m.s. of the fit residuals is 0.0517.
The inset shows the impact of different factors contributing to the
overall fit of Model 4. The LPPL factor is reduced or translated by
the amount of $A$ for a better comparison.}
\end{figure}

\clearpage

\clearpage

\clearpage

\clearpage

\end{document}